\documentclass[iop,apj]{emulateapj}
\slugcomment{{\sc \underline{Accepted for publication in Journal of Geophysical Research - Space Physics:}} April 13, 2017}
\usepackage{amsmath,amssymb,amstext}
\usepackage[breaklinks,colorlinks,citecolor=blue,linkcolor=magenta]{hyperref} 

\usepackage[all]{hypcap} 

\usepackage{natbib,color}
\shorttitle{Gravity Waves and Tides}
\shortauthors{Y\.{I}\u{g}\.{I}t and Medvedev}
\graphicspath{{/Users/erdal/OneDrive/GWtides/Figures/}}

\begin{document}
\title{\textbf{Influence of parameterized small-scale gravity waves on the migrating diurnal
    tide in Earth's thermosphere}}

\author{Erdal Y\.{I}\u{g}\.{I}t\altaffilmark{1}, Alexander S. Medvedev\altaffilmark{2,3}}


\altaffiltext{1}{Department of Physics and Astronomy, Space Weather Laboratory,
 George Mason University, Fairfax, Virginia, USA.}

\altaffiltext{2}{Max Planck Institute for Solar System Research, 
G\"ottingen, Germany.}

 \altaffiltext{3}{Institute of Astrophysics, Georg-August University,
G\"ottingen, Germany.}

\begin{abstract} 
  Effects of subgrid-scale gravity waves (GWs) on the diurnal migrating tides are investigated
  from the mesosphere to the upper thermosphere for September equinox conditions, using a
  general circulation model coupled with the extended spectral nonlinear GW parameterization
  of \citet{Yigit_etal08}.  Simulations with GW effects cut-off above the turbopause and
  included in the entire thermosphere have been conducted. GWs appreciably impact the mean
  circulation and cool the thermosphere down by up to 12-18\%. GWs significantly affect the
  winds modulated by the diurnal migrating tide, in particular in the low-latitude mesosphere
  and lower thermosphere and in the high-latitude thermosphere. These effects depend on the
  mutual correlation of the diurnal phases of the GW forcing and tides: GWs can either enhance
  or reduce the tidal amplitude. In the low-latitude MLT, the correlation between the
  direction of the deposited GW momentum and the tidal phase is positive due to propagation of
  a broad spectrum of GW harmonics through the alternating winds.  In the Northern Hemisphere
  high-latitude thermosphere, GWs act against the tide due to an anti-correlation of tidal
  wind and GW momentum, while in the Southern high-latitudes they
  weakly enhance the tidal amplitude via a combination of a partial correlation of phases and
  GW-induced changes of the circulation. The variable nature of GW effects on the thermal tide
  can be captured in GCMs provided that a GW parameterization (1) considers a broad spectrum
  of harmonics, (2) properly describes their propagation, and (3) correctly accounts for the
  physics of wave breaking/saturation.
\end{abstract}


\section{Introduction}  

Vertically propagating internal gravity waves (GWs) of lower atmospheric origin significantly
affect the dynamics of the middle atmosphere, where their role is widely appreciated
\citep[e.g., see the review of][]{FrittsAlexander03}. Studies over the last decade
demonstrated that these waves can effectively propagate deep into the thermosphere despite the
steeply growing molecular diffusion and thermal conduction \citep{Yigit_etal08, Hickey_etal09,
  Hickey_etal10b,FrittsLund11, GavrilovKshevetskii13, VadasFritts05, Vadas_etal14,
  Heale_etal14}, and considerably influence the state and circulation at altitudes up to the F
region of the ionosphere \citep{Yigit_etal09, Miyoshi_etal14}. The most recent review of these
studies can be found in the works of \citet{YigitMedvedev15,Yigit_etal16b} and in the book by
\citet[Chapter 5,][]{Yigit15}.  In addition to the dynamical forcing, GWs contribute to
heating and cooling of the thermosphere above the turbopause ($\sim$105 km) up to F$_2$ layer
altitudes, reaching peak values of $\sim170 $ K~day$^{-1}$ \citep{YigitMedvedev09}. Since the
thermosphere is strongly susceptible to variations of solar activity, GW propagation and
effects exhibit appreciable changes, as was predicted by general circulation models (GCMs)
\citep{YigitMedvedev10} and recently demonstrated by satellite observations
\citep{Park_etal14, Garcia_etal16}. Transient events in the lower and middle atmosphere, such
as sudden stratospheric warmings, also alter propagation of GWs causing an increase of the GW
activity in the upper thermosphere by more than a factor of three
\citep{YigitMedvedev12,YigitMedvedev16} and generating changes in the thermospheric
circulation up to $\pm50\%$ \citep{Yigit_etal14}. Obviously, in order to achieve a physically
consistent picture of middle and upper atmospheric dynamics, GW effects cannot be ignored, and
any GCM extending to thermospheric heights must properly account for their effects.

Thermal tides are persistent large-scale atmospheric oscillations of the field variables
(temperature, wind, pressure, density) that are generated at various altitudes within the
atmosphere. They are produced by periodic heating due to absorption of incoming solar radiation
\citep{Siebert61}. In the lower atmosphere, they are generated by the absorption of UV
radiation by stratospheric ozone and of IR by tropospheric water vapor \citep{Forbes84}, and by
latent heat release \citep{HaganForbes02, HaganForbes03}.  Thermospheric migrating tide is
excited in situ primarily by photoabsorption of solar radiation in the UV and EUV wavelength
range by molecular oxygen \citep{Fesen_etal86}.

The diurnal westward propagating tide with the zonal wavenumber one (DW1) is a prominent mode
in the mesosphere and the lower thermosphere (MLT) \citep{LiebermanHayes94}. The diurnal
amplitudes in the zonal and meridional wind components demonstrate a distinct semiannual cycle,
and reach maxima during equinoxes \citep{Hays_etal94, Manson_etal02b, Manson_etal02c,
  Wu_etal06, Lieberman_etal10} and minima during solstices \citep{Pancheva_etal02,
  Davis_etal13}.

In the thermosphere, thermal tides emanating from below are supplemented by in-situ generated
solar tides \citep{Hagan_etal01}, which are especially strongly excited in low-latitudes
\citep{Forbes_etal93}. Characterization of tidal signatures of lower atmospheric origin in the
atmosphere above turbopause altitudes and quantification of their dynamical effects on the
thermospheric temperature and mean flow are crucial for better understanding the degree of
vertical coupling \citep{Pancheva_etal06, Knizova_etal15a}. Since the advent of global
satellite observations, propagation of tidal signatures into the thermosphere have been
increasingly studied \citep[e.g.,][]{HuangReber03, Oberheide_etal11, Pancheva_etal12,
  PanchevaMukhtarov12, Hausler_etal15}.  With only a few months of measurements, satellites
can provide good local time and longitudinal coverage.  Analyzing neutral density measurements
by accelerometers onboard the CHAMP (Challenging Minisatellite Payload) and GRACE (Gravity
Recovery and Climate Experiment) satellites, \citet{Forbes_etal09} showed that tides
originating in the troposphere have a significant impact on the MLT, and that the tidal
influences can extend up to the exosphere linking exospheric variability to the variability
near the surface.

Interactions of GWs and thermal tides have been a subject of numerous studies
\citep[e.g.,][]{MiyaharaForbes91, Manson_etal98, Manson_etal02,Manson_etal02b,Manson_etal04,
  Achatz_etal08,Chang_etal08,WatanabeMiyahara09, BeldonMitchel10, SenfAchatz10,LiuA_etal13,
  HealeSnively15}. The majority of these studies focused on
the MLT region, while the influence of GWs on tides in the upper thermosphere received little
attention or no attention at all. Partially, this was caused by a relative scarcity of
observations and by inability of GW parameterizations, which were developed primarily for
middle atmosphere models, to capture GW propagation and effects in whole atmosphere GCMs above
the turbopause. Our study addresses this gap in the knowledge by considering GW effects on the
diurnal migrating tides at altitudes from the tropopause ($\sim$15 km) to the upper
thermosphere ($\sim $300 km), corresponding to the ionospheric F$_2$ layer altitudes.

How do upward propagating thermal tides, in-situ generated thermospheric tides and GWs interact
within the context of vertical coupling?  What are the implications of GW penetration into the
upper atmosphere for the tidal wave structures?  What is needed for GCMs to realistically
capture GW-tide interactions? These questions were the essential motivation for this
paper. Given that internal GWs and tides both grow in amplitude with height, and that they
largely dissipate in the middle and upper atmosphere, interactions of these waves play a
substantial role in the energy and momentum budget in these regions. Various aspects of
GW-tidal interactions have been studied in a number of publications. Using a numerical model of
the diurnal tide coupled with simplified linear GW drag calculations, \citet{MiyaharaForbes91}
showed, considering only harmonics with slow horizontal phase speeds, that GW stress reduces
the tidal amplitudes in the MLT.  With a linear steady-state tidal model complemented by a
linear GW parameterization, \citet{Meyer99a} examined the role of GW effects in the seasonal
variations of the diurnal tides.  Observations of the diurnal tide amplitude and phase
variability with the medium frequency and meteor radars in the MLT revealed a modulation of GW
propagation and momentum fluxes by tides \citep{Nakamura_etal97}. The study of
\citet{Manson_etal02} showed that the tidal winds simulated by a GCM strongly depend on the
utilized GW parameterization. In addition to modeling, interactions between GWs and tides have
been studied with observations \citep[e.g.,][]{Manson_etal98, BeldonMitchel10, LiuA_etal13,
  AgnerLiu15}.
  
In this paper we re-examine the influence of upward propagating small-scale GWs on the
migrating diurnal tide at altitudes from the middle atmosphere to the upper thermosphere.  We
consider the DW1 component because it dominates in the thermosphere, and focus on the
September equinox, when tidal amplitudes are largest. We employ the CMAT2 (Coupled Middle
Atmosphere Thermosphere-2) GCM with the implemented state-of-the-art nonlinear spectral GW
parameterization of \citet{Yigit_etal08}, Global Scale Wave Model
\citep[(GSWM),][]{HaganForbes03} tidal fields and NCEP (National Centers for Environmental
Predictions) reanalysis data at 15 km.
  
The structure of the paper is as follows. Section \ref{sec:model} summarizes the main features
of the CMAT2 GCM, the extended nonlinear GW parameterization of \citet{Yigit_etal08}, and the
setup of the simulations.  Sections~\ref{sec:mean} and \ref{sec:mean-zonal-mean} describe the
modeled mean fields and mean GW effects during the September equinox, respectively.
Simulations of the diurnal migrating tides are presented in section~\ref{sec:solar-tides};
diurnal variations of GW effects are analyzed in section~\ref{sec:gwtide}. Longitudinal
variations are investigated in section~\ref{sec:long-temp-vari}. Section
\ref{sec:discussion} discusses our results in the context of the MLT and upper thermospheric
dynamics. Summary and conclusions are given in section \ref{sec:conc}.

\section{Model Description, Gravity Wave Scheme, and Experiment Design}\label{sec:model}

\begin{table*}[t]
  \caption{Test runs used in the paper. In the cutoff simulation, EXP0, lower atmospheric GWs are not allowed to propagate into the thermosphere above $\sim$105 km, while in the extended simulation, EXP1, GWs propagate into the upper thermosphere.}\label{tb:t1}
  \begin{center}
      \begin{tabular}{cccc}
        \hline\hline
        Model Experiment & Experiment description\\
        \hline
        Cutoff Simulation (EXP0)& Gravity wave effects excluded above 105 km   \\
        Extended Simulation (EXP1) & Gravity wave effects included in the whole atmosphere \\
        \hline
\end{tabular}
\end{center}
\end{table*}

%
%
\begin{figure*}[t]\centering
  \noindent\includegraphics*[width=0.9\textwidth]{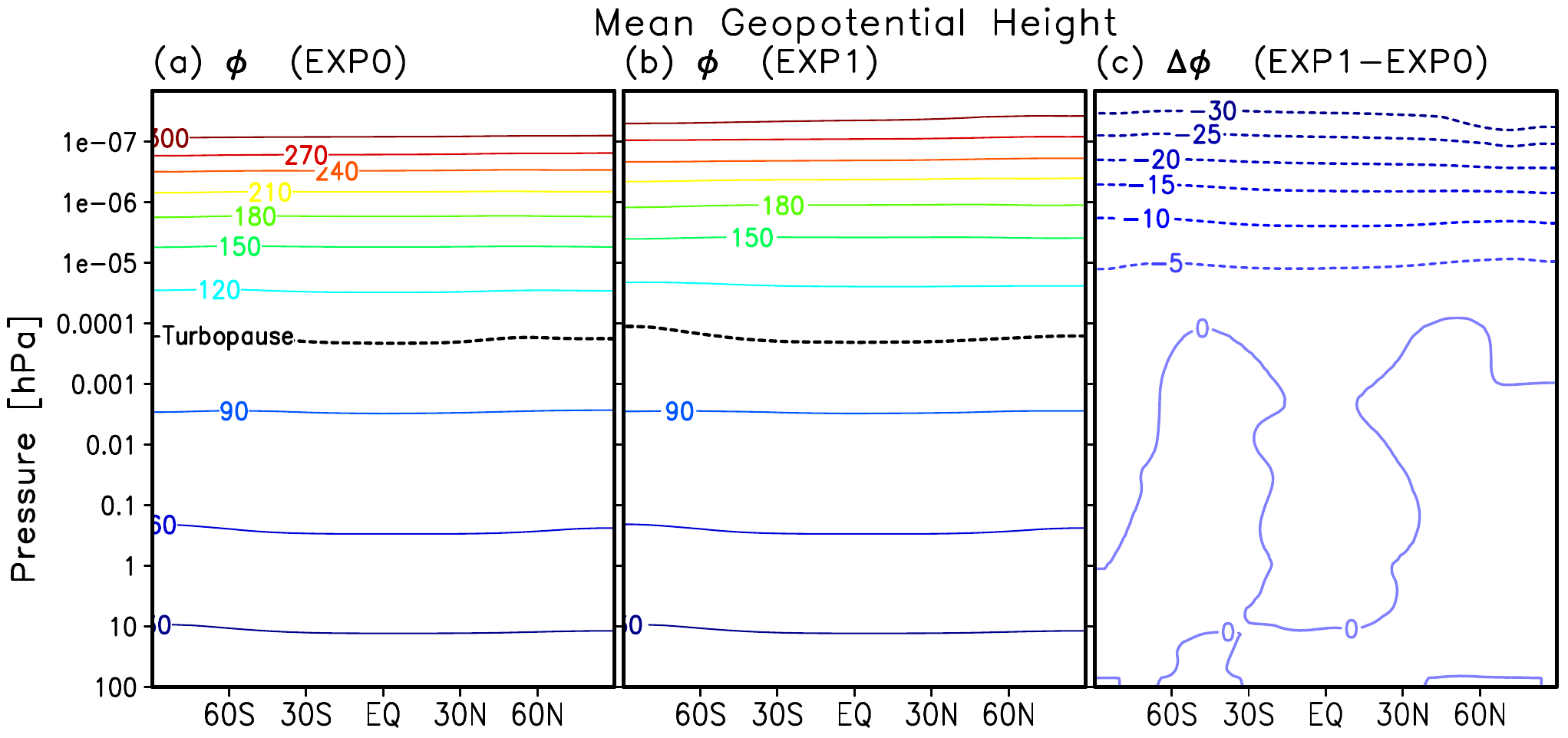}
  \caption{Pressure-latitude distribution of the mean geopotential height [km] in (a) cutoff
    (EXP0) and (b) extended (EXP1) simulation. The difference field (EXP1-EXP0) is shown in
    (c).}
  \label{fig:geopot}
\end{figure*} 

The CMAT2 GCM is a three-dimensional general circulation model that extends from the tropopause
($\sim$15 km, 100 hPa) to the thermosphere to $\sim 250-500$ km ($1.43\times 10^{-8}$ hPa)
depending on the solar and geomagnetic activity. It is a finite difference model that solves
the energy, momentum, and mass conservation equations on a $18^\circ\times 2^\circ$
longitude-latitude grid at 63 fixed pressure levels. At the lower boundary, the model is forced
by the National Centers for Environmental Prediction (NCEP) data, and the solar tides are
specified at the lower boundary from the output of Global Scale Wave Model
\citep[(GSWM),][]{HaganForbes03}. The SOLAR2000 empirical model of \citet{Tobiska_etal00} is
used in order for calculating thermospheric heating, photodissociation, and photoionization due
to absorption of solar X rays, EUV and UV radiation between wavelengths 1.8 and 184 nm.

%
%
\begin{figure*}[t!]\centering
  \vspace{-0cm}
  \noindent\includegraphics[trim=0.cm 1cm 0cm 6cm,clip,width=0.9\textwidth]{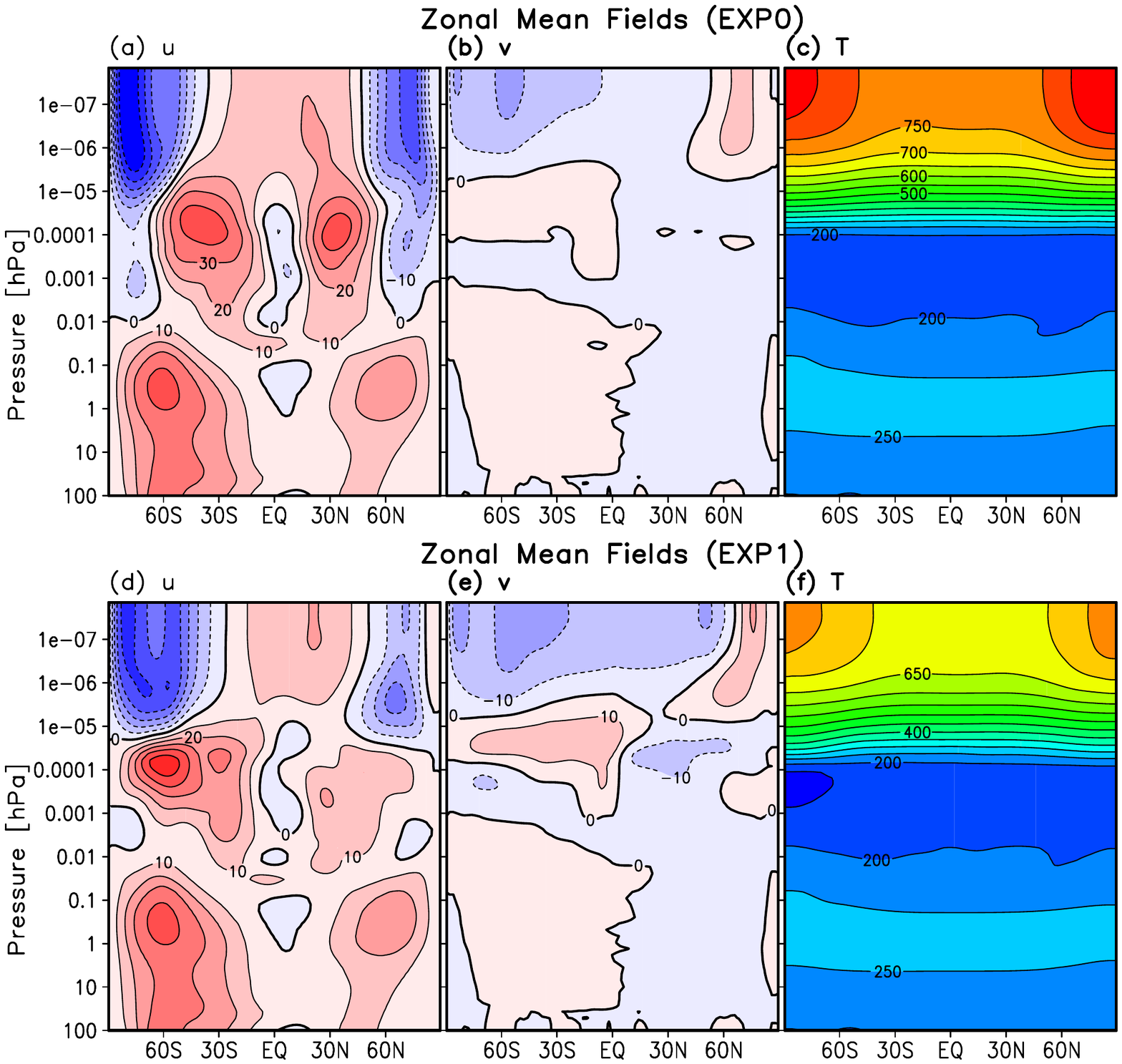}
  \caption{Mean zonal mean atmospheric fields around September equinox for the zonal winds
    $u$, meridional winds $v$, and temperature $T$ from the left to right columns,
    respectively, simulated in the cut-off, i.e., EXP0, (panels a--c, top row) and extended,
    i.e., EXP1, (panels d--f, bottom row) gravity wave simulations. In the zonal wind and
    meridional wind plots, positive values (red shading, solid contour lines) represent
    eastward and northward winds, respectively. The contour intervals are 10 m s$^{-1}$ for
    winds 50 K for the temperature.}
  \label{fig:mean}
\end{figure*}

The model has a conventional (coarse-grid) resolution and cannot self-consistently simulate
small-scale GWs.  Therefore, the influence of subgrid-scale GWs on the large-scale circulation
is represented by the extended spectral nonlinear GW parameterization \citep{Yigit_etal08}. The
development of this scheme has been reviewed in the work by \citet{YigitMedvedev13}. Various
assumptions and limitations of GW parameterizations are discussed in detail in the review of
\citet{YigitMedvedev16}. This scheme accounts for propagation and dissipation of GWs in the
whole atmosphere due to nonlinearity ($\beta_{non}$)
\citep{MedvedevKlaassen95,MedvedevKlaassen00}, molecular diffusion and thermal conduction
($\beta_{mol}$), eddy viscosity ($\beta_{eddy}$), ion drag ($\beta_{ion}$), and radiative
damping ($\beta_{rad}$) in form of Newtonian cooling, which are calculated for each harmonic
$i$ in a broad spectrum of gravity waves ($M=30$). Thus the total dissipation $\beta_{tot}^i$
for a given harmonic is given by
\begin{equation}
  \label{eq:beta}
  \beta_{tot}^i = \sum_n \beta_n^i 
               = \beta_{non}^i + \beta_{mol}^i + \beta_{eddy}^i + \beta_{ion}^i + \beta_{rad}^i, 
\end{equation}
where $n $ denotes the different dissipation terms listed above.  The two major input
parameters are the GW source spectrum, which specifies a distribution of GW
activity in terms of horizontal momentum fluxes $\overline{u^\prime w^\prime_i}$ as a function
of the horizontal phase speed $c_i$ at the lower boundary ($\sim$15 km)
\citep[][\textbf{Figure} 1]{Yigit_etal12b}, where the subscript $i $ denotes a GW harmonic
from the spectrum, and the horizontal wave number $k_H = 2\pi/\lambda_H$. The assumed spectrum
is in a qualitative and quantitative agreement with balloon measurements of GW activity
\citep[\textbf{Figure} 6]{Hertzog_etal08}.  We assume the characteristic scale of
$\lambda_H = 300$ km, which is typical for lower atmospheric GWs to be parameterized
\citep{Samson_etal90}. The GW harmonics are launched in the direction of the local wind at the
lower boundary followed by evaluation of vertical profiles of the momentum fluxes at each grid
point in the model. Although a fixed spectral shape of gravity wave distribution is
  assumed at the source, the actual GW source activity is highly variable as the source winds
  are spatiotemporally variable. The dynamical and thermal effects resulting from the
divergences of the momentum fluxes are calculated as described in the works by
\citet{Yigit_etal09} and \citet{YigitMedvedev09} and interactively applied to the simulated
(large-scale) fields.  The dynamical effects include the zonal ($a_x $) and meridional
($a_y $) GW drag given by
\begin{equation}
  \label{eq:ax}
  a_x = -\frac{1}{\rho} \frac{\partial (\rho\overline{u^\prime w^\prime})}{\partial x},
\qquad
  a_y = -\frac{1}{\rho} \frac{\partial (\rho\overline{v^\prime w^\prime})}{\partial y},
\end{equation}
where $\rho $ is the background neutral density. The thermal effects $Q $ incorporate the
total resulting effects from irreversible heating $Q_{irr} $ and differential heating/cooling
$Q_{dif}$ \citep[Eqn.  1]{YigitMedvedev09}.

%
%
\begin{figure*}[t!]\centering
  \vspace{-0cm}\includegraphics[trim=0.cm 1cm 0cm 15cm,clip,width=0.95\textwidth]{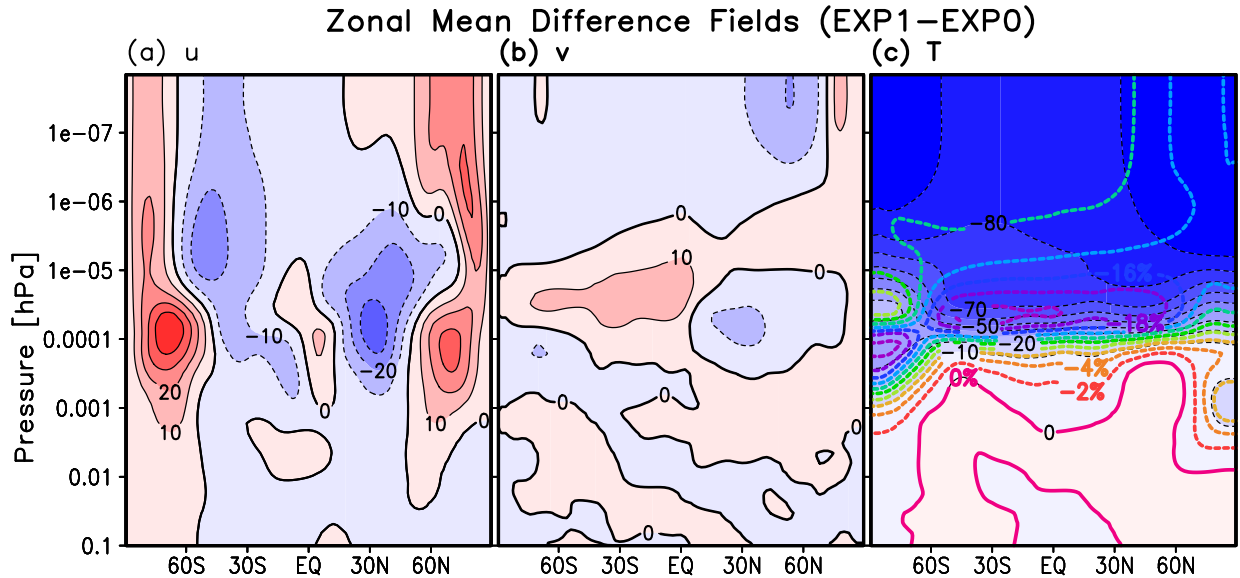}
  \caption{Mean atmospheric difference fields obtained by subtracting the cutoff simulation
    results from the extended simulation (EXP1-EXP0). }
  \label{fig:mean_dif}
\end{figure*}

The extended scheme has been validated in the works by \citet{Yigit_etal09} and
\citet{YigitMedvedev09} with respect to MSIS and HWM (Horizontal Wind Model), two established
empirical models of the thermosphere. The scheme was numerously applied in studies of
atmospheric dynamics \citep{Yigit_etal09, Yigit_etal12b, Yigit_etal14, YigitMedvedev09,
  YigitMedvedev10, YigitMedvedev12}. It is also being used for Martian GCMs
\citep{Medvedev_etal11a, Medvedev_etal11b, Medvedev_etal13, Medvedev_etal15, Medvedev_etal16,
  MedvedevYigit12, Yigit_etal15a, Yigit_etal15b}, and has been most recently successfully
applied for interpretation of MAVEN (Mars Atmosphere Volatile EvolutioN) observations of
high-altitude GWs in the thermosphere of Mars \citep{Yigit_etal15b}.

The model is run for low solar and geomagnetic activity conditions ($F_{10.7}$ = 80 W m$^{-2}$
Hz$^{-1}$, $K_p$ = 2+) for about a half year, and data are output for every four hours around
the September equinox over the period 19--25 September (i.e., about one week centered around
the Northern Hemisphere autumnal equinox). This excludes the well known influence of
high solar activity on GWs \citep{YigitMedvedev10,Laskar_etal15} and larger-scale circulation 
(including in-situ excited tides) in the thermosphere, but still captures the mechanism of GW 
influence on the tides, which is of interest here. September zonal mean fields have been 
validated with respect to the HWM previously \citep{Yigit_etal12b}. In this study, in order to 
quantify GW effects on the migrating diurnal tide, we use simulation data produced by running 
the model in two modes: (1) with out and (2) with GW propagation into the thermosphere above 
the turbopause ($\sim$105 km), referred to as the ``cutoff" (EXP0) and the ``extended" (EXP1)
simulations, respectively, as summarized in Table \ref{tb:t1}. 

In practical terms, the notions of ``cutoff" and ``extended" simulations refer to the
methodologies of excluding and including, respectively, the thermospheric effects of gravity
waves of lower atmospheric origin, as calculated by the extended nonlinear GW parameterization
\citep{Yigit_etal08}. By default, this scheme calculates GW effects in the entire atmosphere,
which corresponds to the ``extended" case (``EXP1") of the GCM simulation. In order to assess
the significance GW propagation into the thermosphere, we additionally perform the cutoff
simulation, which is designed to incorporate GW effects calculated by the GW scheme only from
the lower atmosphere up to the turbopause.

In both cases, the GW effects are analyzed, and the
modeled zonal ($u$) and meridional winds ($v$) as well as neutral temperature ($T$) fields are
Fourier decomposed in order to obtain diurnal migrating tide amplitudes. We compare the cutoff
and the extended simulation results at fixed pressure levels.

To provide a further perspective on the thermospheric extension of the model levels, we
plotted the zonal mean geopotential height $Z_\Phi$ (in km) averaged over the chosen time
interval in the entire model domain in \textbf{Figure}~\ref{fig:geopot} for (a) EXP0 (cutoff),
(b) EXP1 (extended), and (c) the difference between the extended and the cutoff simulations,
i.e., EXP1--EXP0. Overall, the model covers altitudes from $Z_\Phi = 15$ km to about 300 km
under the assumed low solar and geomagnetic conditions.  The turbopause reference height of
$\sim$105 km is close to the $10^{-4} $ hPa pressure level and varies little between the two
simulations. It is seen from \textbf{Figure}~\ref{fig:geopot}c that, below the turbopause,
geopotential heights are nearly identical in both runs for all model pressure levels. Above
it, accounting for the thermospheric effects of GWs leads to contraction of the thermosphere,
i.e., thermospheric layer thickness reduces, and to drop of the geopotential heights of the
fixed pressure levels relative to the cutoff simulation.  The difference $\Delta Z_{\Phi}$ at
fixed pressure levels due to GW effects vary from --5 km in the lower thermosphere to --30 km
in the upper thermosphere.

%
%
\begin{figure*}[t!]\centering
\includegraphics[trim=0.cm 1cm 0cm 5.cm,clip,width=0.9\textwidth]{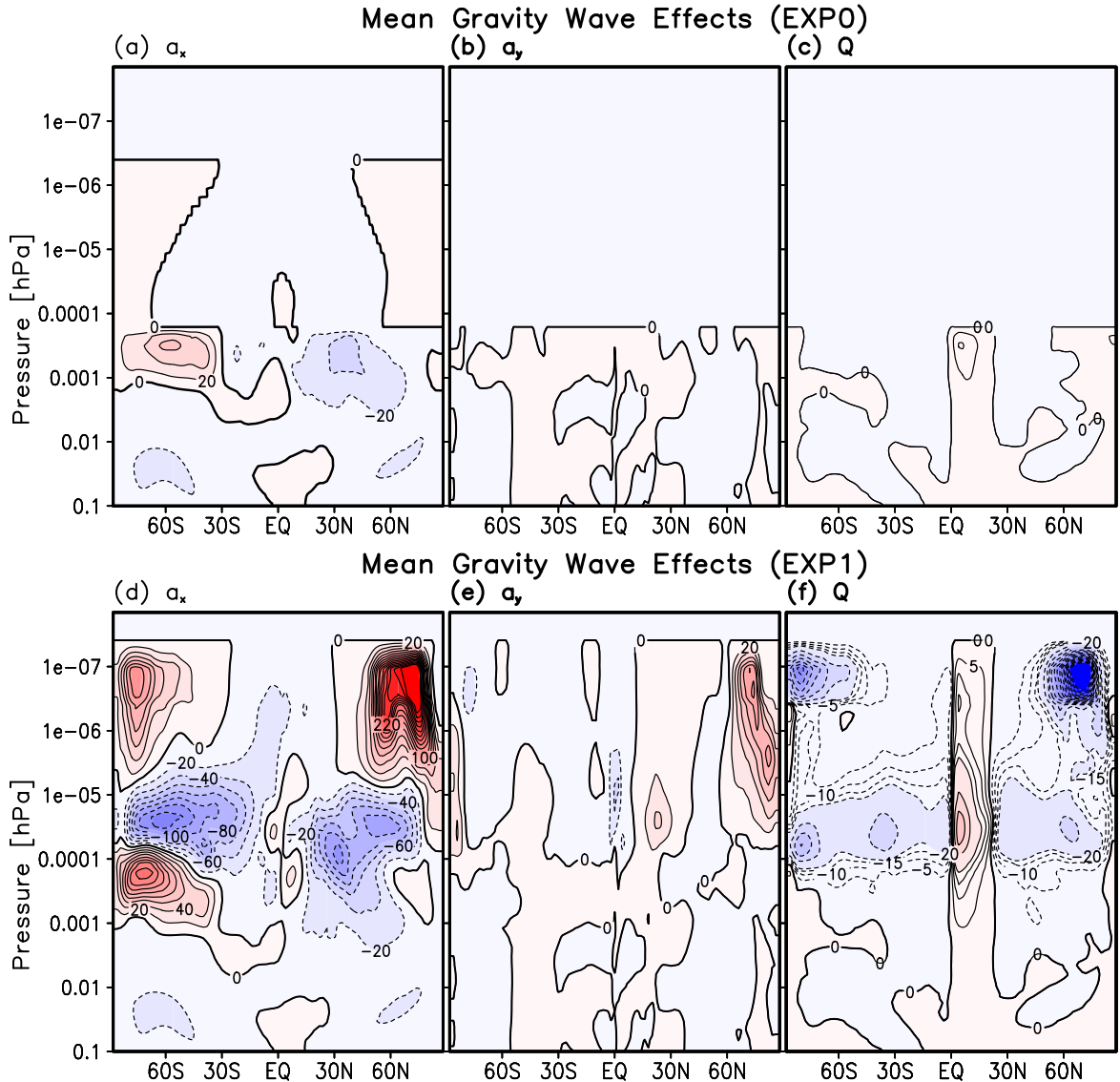}
  \caption{Same as Figure \ref{fig:mean} but for the GW dynamical ($a_x, a_y$;
    acceleration/deceleration) and $Q $; thermal (heating/cooling) effects in m s$^{-1}$
    day$^{-1}$ and K day$^{-1}$, respectively. Contour intervals are 20 m s$^{-1}$ day$^{-1}$
    and 20 K~day$^{-1}$. For $Q $, $\pm 5, \pm 10, and \pm 15$ K~day$^{-1}$ contours have been added.}
  \label{fig:mean_gw}
\end{figure*}

\section{Zonal Mean Fields}\label{sec:mean}

%
%
\begin{figure*}[t!]\centering\vspace{-0cm}
 \includegraphics[trim=0.cm 1cm 0cm 6.cm,clip,width=0.9\textwidth]{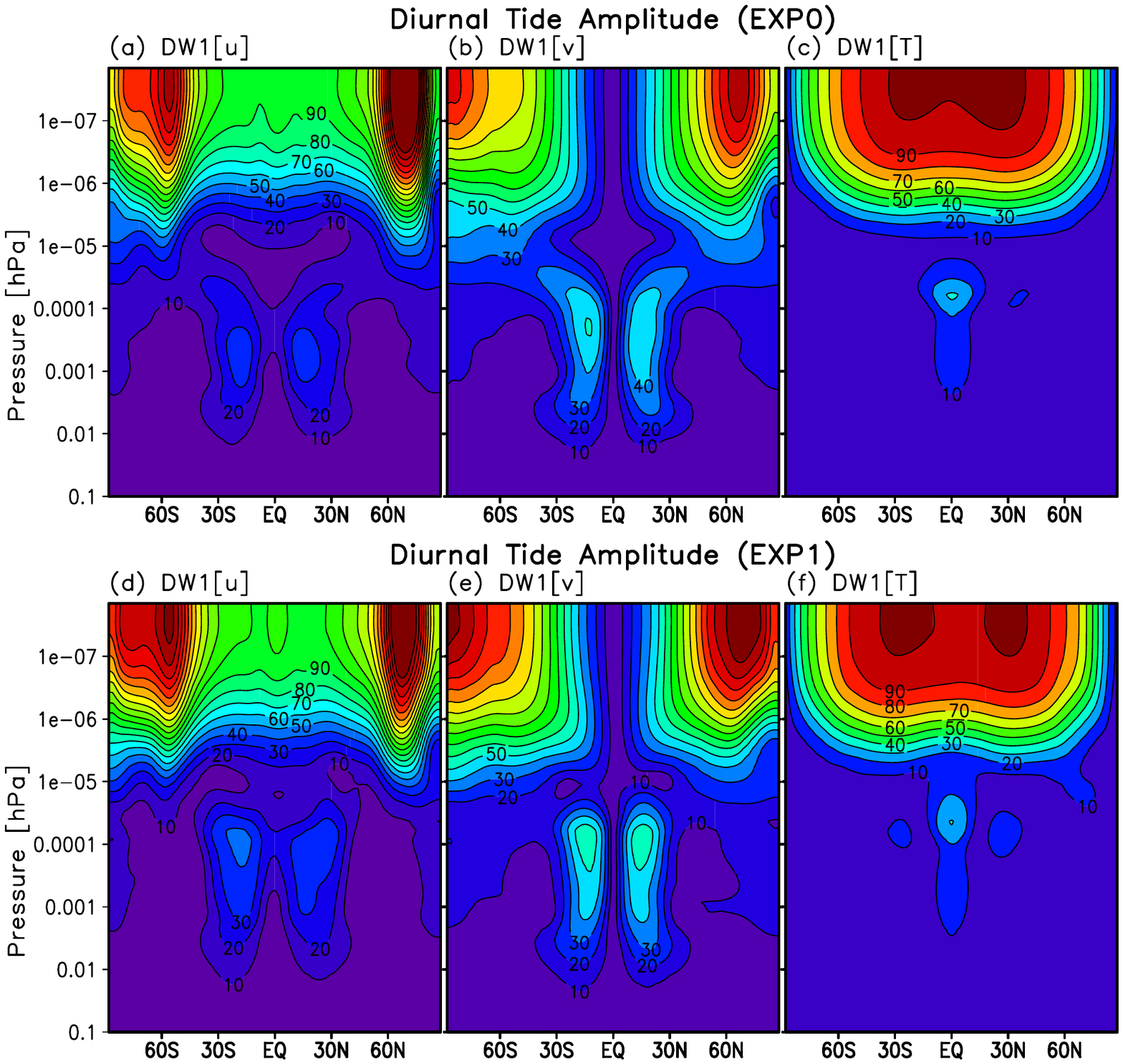}
  \caption{Altitude (pressure)-latitude distributions of the amplitude of the diurnal
    migrating tide (DW1) for the zonal wind [m s$^{-1}$], meridional wind [m s$^{-1}$], and
    temperature [K] shown from left to right in the cut-off (panels a--c) and the extended
    (panels d--f) simulations.  }
  \label{fig:tideamp}
\end{figure*}

We first analyze the altitude-latitude cross-sections of the zonal mean fields for the
September equinox and the corresponding differences between the extended and the
  cutoff simulations.  The columns from left to right in \textbf{Figure}~\ref{fig:mean}
present the simulated mean zonal wind ($\bar{u}$), meridional wind ($\bar{v}$),
and neutral temperature ($\bar{T}$), correspondingly, for the cutoff (EXP0, top row) and
extended (EXP1, bottom row) runs.  Red shading and solid lines denote westerly (eastward)
winds, while blue shading and dashed lines are for easterly winds (westward) in units of
m~s$^{-1}$.

During the equinox, westerly zonal winds are predominant in the lower and middle atmosphere and
are relatively weak (peak values are up to 40 m~s$^{-1}$). Their extension into the upper
thermosphere is modified primarily by Coriolis force, ion drag and high-latitude energy
sources (e.g., Joule heating and particle precipitation) in addition to the rapidly increasing
molecular diffusion. The equinoctial meridional circulation is also relatively weak (compared
to the solstitial one), and the mean neutral temperature increases up to $\sim$750--900 K in
the upper thermosphere.

Closer inspection of the mean fields reveals the major differences above 100 km between the
two simulations, in particular for the zonal mean wind at middle- and high-latitudes, which we
consider in detail in \textbf{Figure}~\ref{fig:mean_dif} by evaluating the corresponding
differences $\Delta \bar{u} $, $\Delta \bar{v} $, and $\Delta \bar{T}$ at fixed pressure
levels.  It is seen that accounting for GWs in the whole atmosphere provides more westerly
(positive, red) momentum to the zonal wind in the high-latitude thermosphere, and more
easterly momentum (blue shaded) in the middle-latitude thermosphere in both hemispheres. The
strongest attenuation of the easterly wind takes place in the high-latitude F$_2$ region
altitudes of the Northern Hemisphere (panel a).  It leads to a weakening of the poleward
meridional circulation in both hemispheres (panel b).  One (already
mentioned) remarkable thermal effect of GWs is cooling the lower thermosphere down by about 70
K ($18\%$) at low- and middle-latitudes.  In the upper thermosphere, the temperature
difference is even larger (80--90 K colder), although somewhat smaller in relative units:
$ > 12-14\%$. These GW-induced changes in the background conditions should be kept in mind
when analyzing vertical propagation of GWs and the tide.

%
%
\begin{figure*}[t]\centering
\includegraphics[trim=0.cm 1cm 0cm 15cm,clip,width=0.9\textwidth]{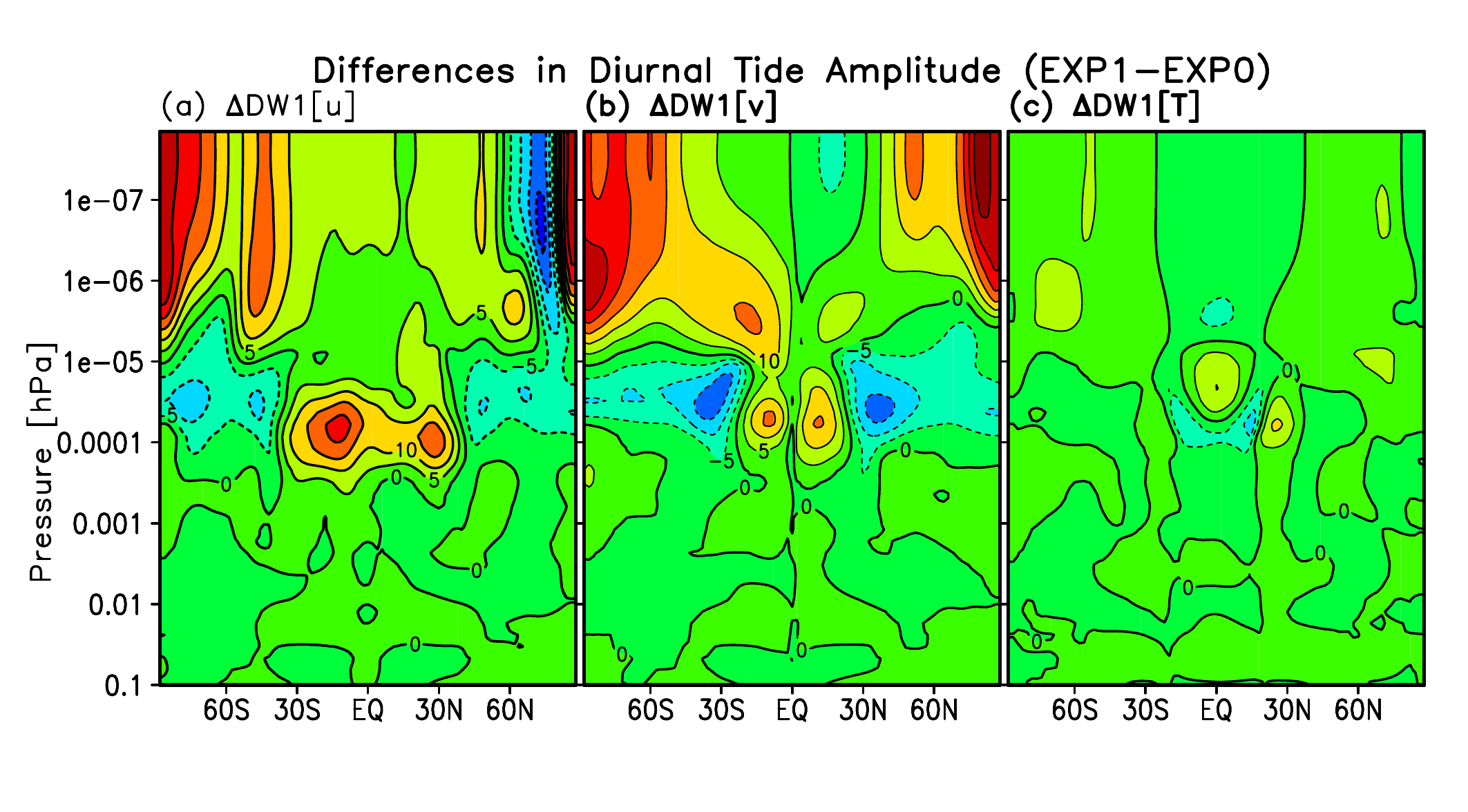}
\caption{Differences in the diurnal tidal (DW1) amplitudes of $u$, $v$, and $T$ between the
  extended and the cutoff simulations, i.e., EXP1-EXP0. Solid/dashed contours represent
  positive/negative differences. The contour intervals are 5 m s$^{-1}$ for the winds and 5 K
  for temperature.}
  \label{fig:tideamp_dif}
   
\end{figure*}

\section{Zonal Mean Gravity Wave Effects} \label{sec:mean-zonal-mean} 

%
%
\begin{figure*}[t!]\centering
  \includegraphics[trim=0.cm 1cm 0cm 4.5cm,clip,width=0.9\textwidth]{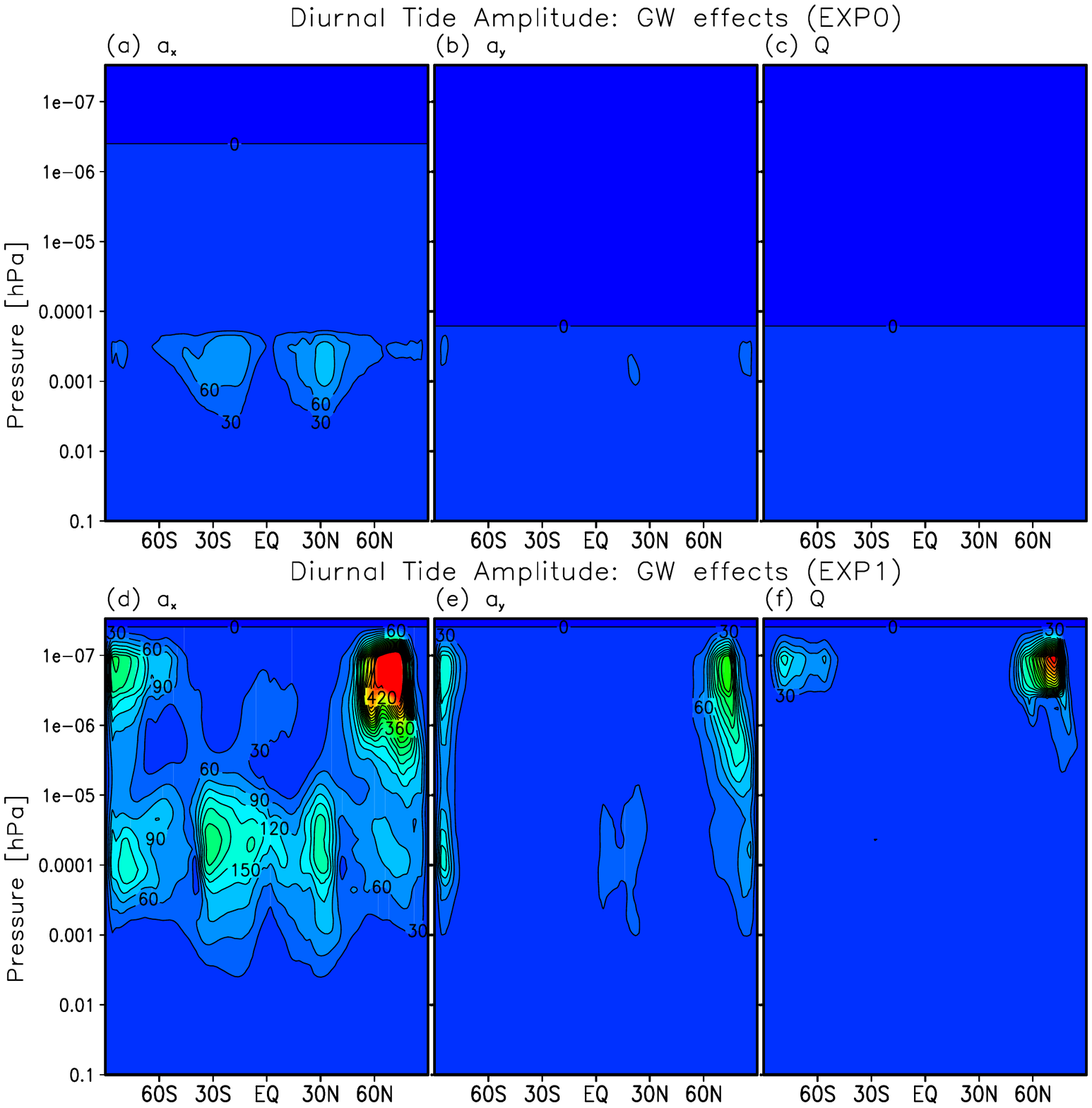}
  \caption{Diurnal tidal amplitudes as in Figure \ref{fig:tideamp} but for the GW zonal drag
    $a_x$, meridional drag $a_y $, and the total heating/cooling $Q $. The controur intervals
    are 30 m s$^{-1}$ day$^{-1}$ for drag and 30 K for heating/cooling.}
  \label{fig:diurnal_gw}
\end{figure*}

The only systematic difference between the two simulations, EXP0 and EXP1, is the inclusion of
the parameterized GW effects above $\sim $105 km. The parameterization scheme allows for
quantifying these effects in order to gain further insight into the nature of changes in the
modeled mean fields. \textbf{Figure}~\ref{fig:mean_gw} presents the zonal mean ($\bar{a}_x$,
panels a,d) and meridional ($\bar{a}_y$, panels b,e) GW forcing (``GW drag") and the total
GW-induced heating/cooling rates ($\bar{Q}$, panels c,f) for both simulations in a manner
similar to that in \textbf{Figure}~\ref{fig:mean}. The cutoff simulation illustrates the GW
effects in the MLT on the mean zonal circulation: a weak westward drag of around --20
m~s$^{-1}$~day$^{-1}$ in the stratosphere and +60 and --40 m~s$^{-1}$~day$^{-1}$ in the MLT of
the Southern and Northern Hemispheres, respectively. GWs impose a negligible mean meridional
forcing and heating/cooling in the middle atmosphere.  However, when GWs are allowed to
propagate in the simulations beyond the turbopause, they produce substantial dynamical
effects, primarily the eastward zonal GW drag in the lower and upper thermosphere and westward
GW drag of more than --120 m~s$^{-1}$~day$^{-1}$ in the middle-latitude middle
thermosphere. Note also that GW drag increases in magnitude in EXP1 even below 105 km as a
consequence of the downward control. The latter example emphasizes the
importance of a correct accounting for GW physics in the thermosphere even if the area of
interest is the MLT only.

The zonally averaged meridional GW forcing of $\sim\pm 20$ m~s$^{-1}$~day$^{-1}$ around the
equatorial lower thermosphere and 80 m~s$^{-1}$~day$^{-1}$ in the high-latitude Northern
thermosphere is expectedly smaller than that in the zonal direction. In the lower thermosphere,
GWs provide cooling at middle- and high-latitudes ($\sim -40$ K~day$^{-1}$) and heating
in a narrow latitude band near the equator with 60 K~day$^{-1}$. The thermal effects in
the upper thermosphere are much larger. They include an intensive cooling at high-latitudes
exceeding --100 K~day$^{-1}$ and $\sim -260$ K~day$^{-1}$ in the Southern and Northern
Hemispheres, respectively. One may notice a certain degree of symmetry of GW-induced effects
with respect to the equator during the equinox, although there are some noticeable
interhemispheric differences in the magnitude of the effects due the interplay of background
atmospheric wave filtering and dissipation.

The GW-induced differences in the simulated mean fields presented in
\textbf{Figure}~\ref{fig:mean_dif} can now be interpreted better together with the GW dynamical
and thermal effects explicitly shown in \textbf{Figure}~\ref{fig:mean_gw}.  Although the
distributions of GW forcing and the induced changes generally agree, they do not fully
coincide. This phenomenon illustrates that atmospheric dynamics is strongly nonlinear,
interactions of GWs with the mean flow should not be viewed overly simplistic, and that
employed GW parameterizations must capture the interactions with the larger-scale atmosphere,
i.e., be physically based rather than mechanistic.

%
%
\begin{figure*}[t]\centering
  \includegraphics[width=0.9\textwidth]{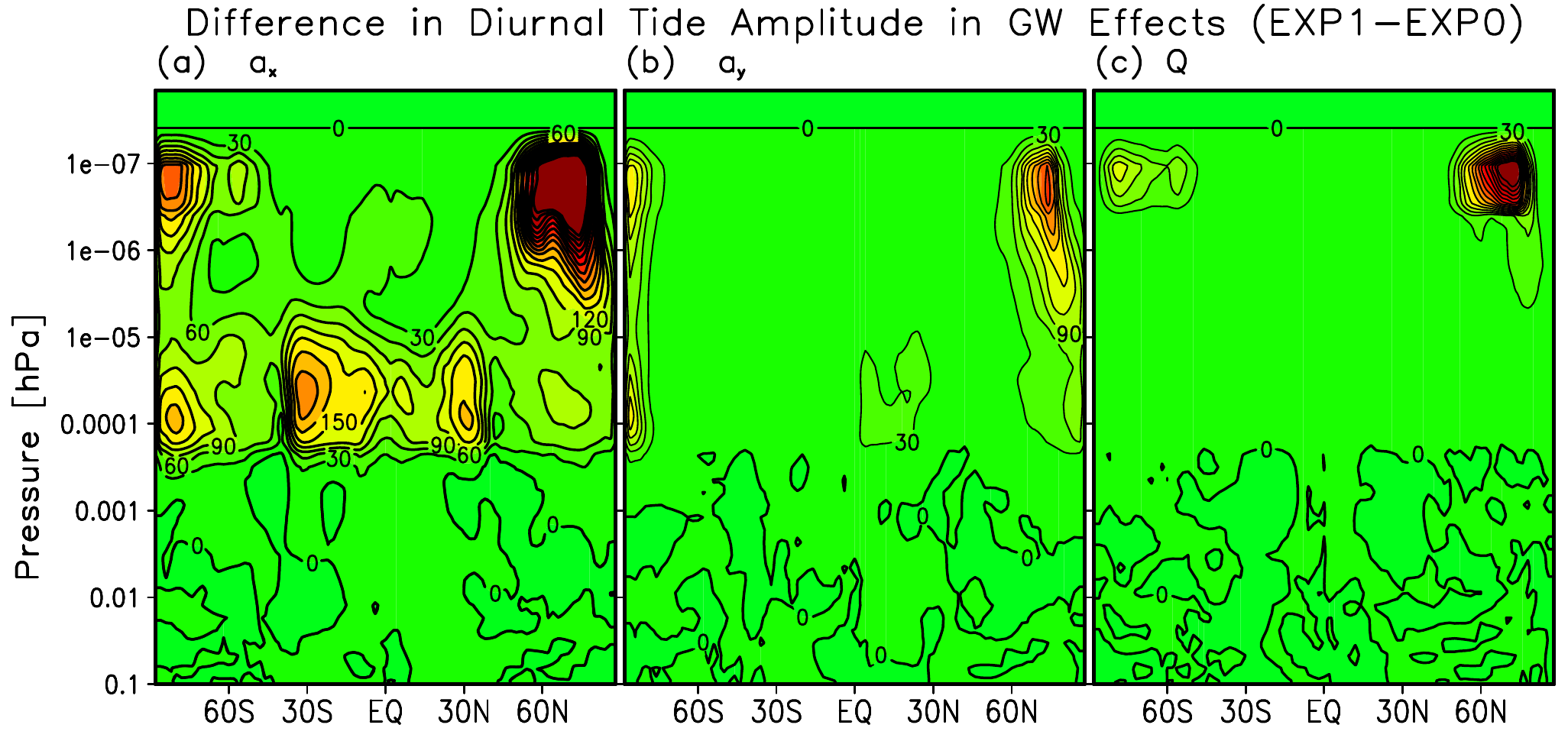}
  \caption{Difference between the extended and the cutoff simulations in the diurnal tidal
    amplitudes for $a_x, a_y $, and $Q $. The controur intervals
    are 30 m s$^{-1}$ day$^{-1}$ for drag and 30 K for heating/cooling.}
  \label{fig:diurnal_gw_dif}
\end{figure*}

\section{Diurnal Migrating Solar Tide}\label{sec:solar-tides}

Model simulations considered above demonstrated that propagation and dissipation of GWs in the
thermosphere above the turbopause result in significant changes in the general circulation and
mean thermal structure of the atmosphere. This finding brings in the main question of this
paper: To what extent do GW effects impact the diurnal tide in the middle and upper
atmosphere?  For this, we next consider the tidal fields in winds and neutral
temperature. They represent composites over the model 19-25 September and, thus, offset a
day-to-day tidal variability \citep{Liu14book}.

\textbf{Figure} \ref{fig:tideamp} shows the diurnal tide amplitudes for the same components
($u$, $v$, and $T$) as the mean fields presented in Figure~\ref{fig:mean} for the both
runs. There are two disparate maxima. The first one is in the low-latitude MLT, and is created
by the tides generated in the lower atmosphere and propagating upward. The much larger
(velocity amplitudes are more than 150 m~s$^{-1}$) maximum in the upper thermosphere (above
$\sim$150 km) is due to the tide generated in-situ by absorption of solar EUV radiation and
ion-neutral coupling \citep{Fesen_etal86}. The structure and amplitude of the diurnal migrating
tide in the MLT are well documented with observations
\citep[e.g.,][]{McLandress_etal96b,Manson_etal02c, Wu_etal06, Lieberman_etal10, AgnerLiu15}.
The simulated morphologies of the tidal amplitudes in the MLT are in a good qualitative
agreement with previous observations. Namely, the amplitudes of the zonal and meridional wind
components peak around 20$^\circ $N and $20^\circ $S at around 100 km with the amplitude of
meridional wind variations exceeding that for the zonal component, and have a distinctive
``butterfly" pattern consistent with the TIDI (TIMED Doppler Interferometer) measurements
\citep{Wu_etal06,Wu_etal08}.  The amplitude of temperature tidal perturbations has the maximum
(15-20 K) over the equator in the MLT, and at low-latitudes in the upper thermosphere
($\sim$100 K).

There are some differences in the magnitude and structure of the tide between the two
simulations. They are plotted in \textbf{Figure}~\ref{fig:tideamp_dif}.  Substantial
differences in the amplitudes of the zonal and meridional winds are seen above the turbopause
and, to a minor extent, in the MLT. GW-induced changes in the temperature component of the
diurnal tide amplitude are mainly confined to the equatorial region of the MLT.  Accounting for
GW effects in the simulations changes diurnal fluctuations of temperature by $\pm$10 K in the
lower thermosphere, or up to 30\% of the tide amplitude.  For the zonal and meridional winds,
GWs facilitate an increase of tidal amplitudes in the low-latitude MLT, and decrease them at
higher latitudes slightly higher (around 120--140 km). The amplitude of zonal wind tidal
oscillations increases by up to 20 m~s$^{-1}$ between $\pm 30^\circ$ and decreases by up
to $10$ m~s$^{-1}$ poleward of $\pm 45^\circ$.

%
%
\begin{figure*}[t!]\centering
  \includegraphics[trim=0.8cm 1cm 1.3cm 12.3cm,clip,width=0.9\textwidth] {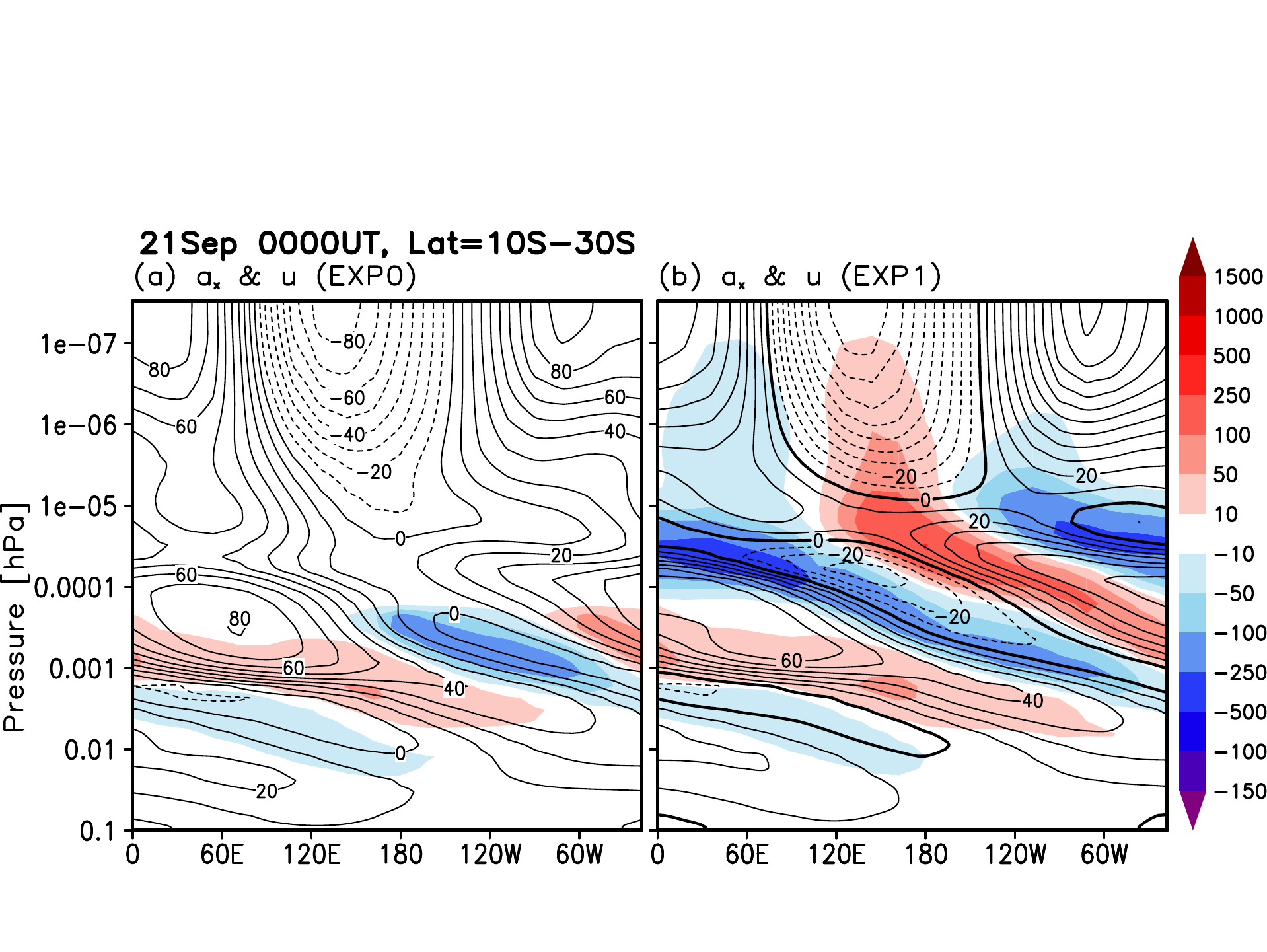}%
  \caption{Pressure-longitude distribution of Southern Hemisphere low-latitude average
    ($10^\circ-30^\circ $S) of the zonal wind (contour) [m s$^{-1}$] and zonal GW drag (shaded)
    [m s$^{-1}$ day$^{-1}$] at a representative universal time (0000 UT) on 21 Sept 0000 UT in
    the cutoff (EXP0) and the extended simulation (EXP1). The zonal wind contour intervals are
    10 m s$^{-1}$.}
  \label{fig:u_ax_wind_p_lon_lowlat}
\end{figure*}

In the thermosphere above $\sim$150 km, the inclusion of parameterized GW effects resulted in
the tidal wind amplitudes increase by up to 25 m~s$^{-1}$, with the exception of zonal winds at
high latitudes of the Northern Hemisphere. There, the amplitudes decreased by up to $\sim$--20
m~s$^{-1}$. The tidal amplitude change in temperature is $\sim$5 K in the high-latitudes, and
about $10$ K in the low-latitudes. Overall, the simulations show where and how GWs impact the
diurnal tide most significantly. In the horizontal wind fields, the effect of the parameterized
GWs occurs nearly at all latitudes in the MLT and, predominantly, at middle- and high-latitudes
in the upper thermosphere, where tidal variations of wind are largest. Similarly,
the temperature component of the diurnal tide is affected, primarily, in the equatorial
thermosphere, where the diurnal temperature fluctuations are large. 

We next embark on further details of how GWs can produce the changes in amplitude and structure
of the DW1 tide by analyzing diurnal variations of the GW drag.  Previous studies that were
concerned mainly with the MLT have indicated that the effects of parameterized GWs depend on
the relative phase between the GW forcing and the tides \citep{LiuA_etal13}.

\section{Diurnal Variation of Gravity Wave Effects}\label{sec:gwtide}

%
%
\begin{figure*}[t!]\centering
  \includegraphics[trim=0.9cm 1cm 1.cm 4cm,clip,width=0.9\textwidth]{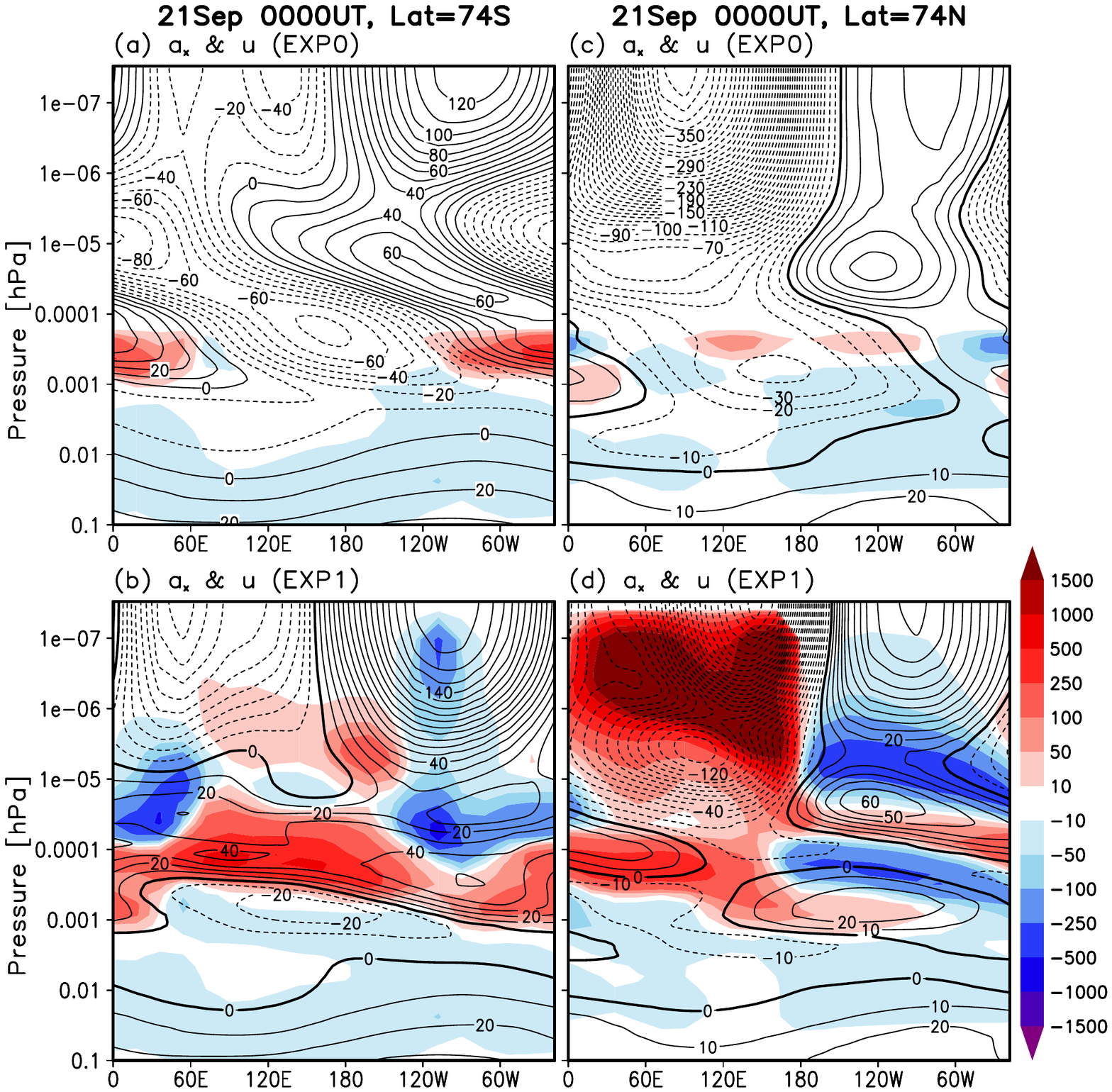}%
  \caption{Same as Figure \ref{fig:u_ax_wind_p_lon_lowlat} but for the Southern Hemisphere
    ($74^\circ $S) and Northern Hemisphere ($74^\circ $N) high-latitudes.}
  \label{fig:u_ax_wind_p_lon_highlat}
\end{figure*}

%
%
\begin{figure*}[t!]\centering
  \includegraphics[trim=0.8cm 1cm 1.3cm 12.3cm,clip,width=0.9\textwidth]{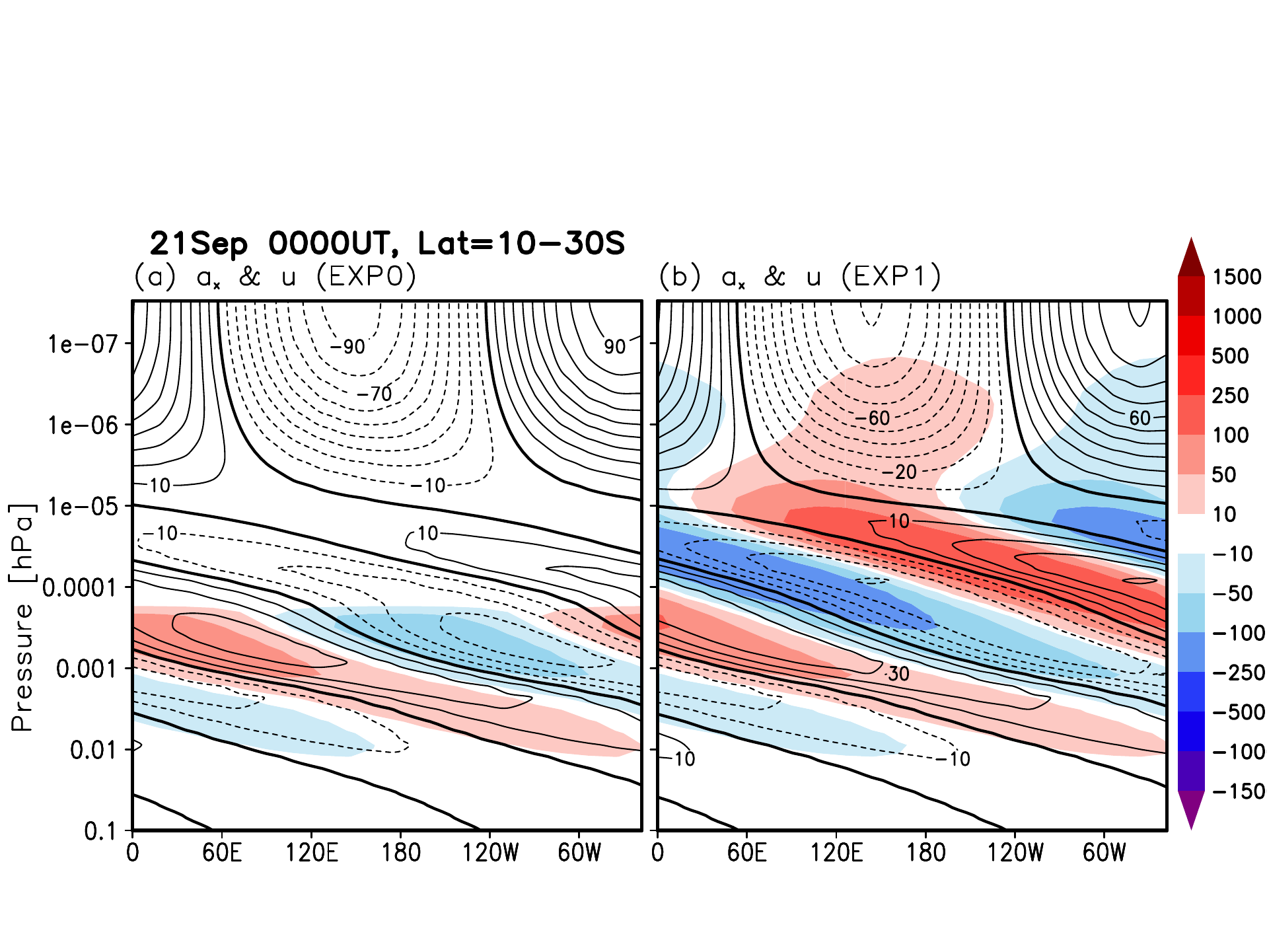}%
  \caption{Diurnal component of the zonal wind and the zonal GW drag presented in Figure
    \ref{fig:u_ax_wind_p_lon_lowlat} at the Southern Hemisphere low-latitudes on 21 Sept 0000
    UT.}
  \label{fig:u_ax_wind_p_lon_lowlat_diurnal}
\end{figure*}

We next investigate diurnal variations of GW effects. \textbf{Figure} \ref{fig:diurnal_gw}
presents the diurnal tidal amplitudes of zonal GW drag, meridional GW drag, and the total GW
heating/cooling rate, for the cut-off and extended simulations, presented in a similar way as
in the preceeding figures. Overall, the largest diurnal variations are seen in the zonal GW
drag.  In the cut-off simulation, these variations are confined to the low-latitude MLT with
peak values of $\sim$90 m~s$^{-1}$~day$^{-1}$ centered around $\sim$$\pm 30^\circ $ latitude. The
meridional GW drag and GW heating/cooling rates show very weak diurnal variations. In the
extended simulation, GW diurnal variations are more remarkable in all GW parameters. The peak
values of the zonal GW drag variations are situated around the low-latitude MLT, have a similar
structure as in the cut-off simulation, but much stronger amplitude of up to 210 m s$^{-1}$
day$^{-1}$, extending to higher altitudes in the thermosphere. Overall, some degree of
hemispheric asymmetry is seen in the diurnal tidal signature in the zonal GW drag.

At higher levels in the thermosphere, the zonal GW drag demonstrates prominent diurnal
variations at high-latitudes with peak values of $\sim $200 m s$^{-1}$ day$^{-1}$ in the
Southern Hemisphere and several hundred m~s$^{-1}$~day$^{-1}$ in the Northern
high-latitudes. The maximum of diurnal variations of the meridional GW drag and total GW
heating/cooling peak in the high-latitudes thermosphere of the Northern Hemisphere.

Analyzing the difference in the GW diurnal variations in \textbf{Figure}
\ref{fig:diurnal_gw_dif} between the cutoff and the extended simulations demonstrates that the
GW diurnal variations are large in the extended simulation in all regions. This suggests that
GW propagation into the thermosphere is strongly modulated by the diurnal tide, and that
GW-tide interactions and nonlinear feedback would not be captured if GWs are not accounted for
in the whole atmosphere system. The diurnal tidal variations in the MLT extend higher up in the
lower thermosphere in the extended simulation. It is noteworthy that the lower thermospheric GW
effects produce changes below the turbopause, as indicated by the nonzero difference in the
diurnal tidal amplitude in $a_x $ (panel a). The simulations also highlight that the zonal GW
drag experiences significant diurnal variations in the lower thermosphere above the
turbopause. Comparison of differences in the diurnal tidal amplitudes in the fields
(\textbf{Figure}~\ref{fig:tideamp_dif}) with the differences in the diurnal amplitudes of GW
effects (\textbf{Figure} \ref{fig:diurnal_gw_dif}) provides further insight into interactions of GWs and tides. In the low-latitude lower
thermosphere at around $\pm 30^\circ $, the enhancement of the GW diurnal amplitude coincides
very well with the enhancement of tidal variations, in particular, of the zonal wind, when GW
effects are included in the whole atmosphere. In the upper thermosphere, strongest diurnal
variations of the GW drag occur in the regions where either the enhancement of diurnal
oscillations of wind takes place, i.e., in the high-latitude Southern Hemisphere, or the
amplitude is strongly reduced (in the high-latitude Northern Hemisphere). We next focus on
these regions where largest diurnal variations of GW effects lead opposite consequences for the
tidal amplitudes.

\section{Longitudinal Variations}\label{sec:long-temp-vari}

Next, we assess the altitude-longitude distributions of the zonal winds and zonal component of
GW drag during the September equinox at a representative universal time (00:00 UT) in order to
gain further insight into GW-tidal interactions.
\textbf{Figure}~\ref{fig:u_ax_wind_p_lon_lowlat} presents the Southern Hemisphere zonal wind
fields (contour) and the zonal GW drag (color shades) for the cutoff (EXP0, top panel) and the
extended (EXP1, bottom panel) simulations averaged over low-latitudes 10$^\circ - 30^\circ$S.
It is seen that, in the low-latitude MLT, the directions of the GW drag and the wind remarkably
coincide, that is, GWs accelerate the local winds rather than ``drag" them.  In the EXP1
simulation, this directional correlation of the wind and GW forcing extends higher into the
thermosphere up to $10^{-5}$ hPa ($\sim$140 km). Further above this height, the in-situ
generated tide takes over the one propagating from below, and the effect of GWs changes
accordingly -- GWs act to weakly decelerate the zonal flow in the middle and upper thermosphere
at low-latitudes.

%
%
\begin{figure*}[t!]\centering
  \includegraphics[trim=0.9cm 1cm 1.cm 4cm,clip,width=0.95\textwidth]{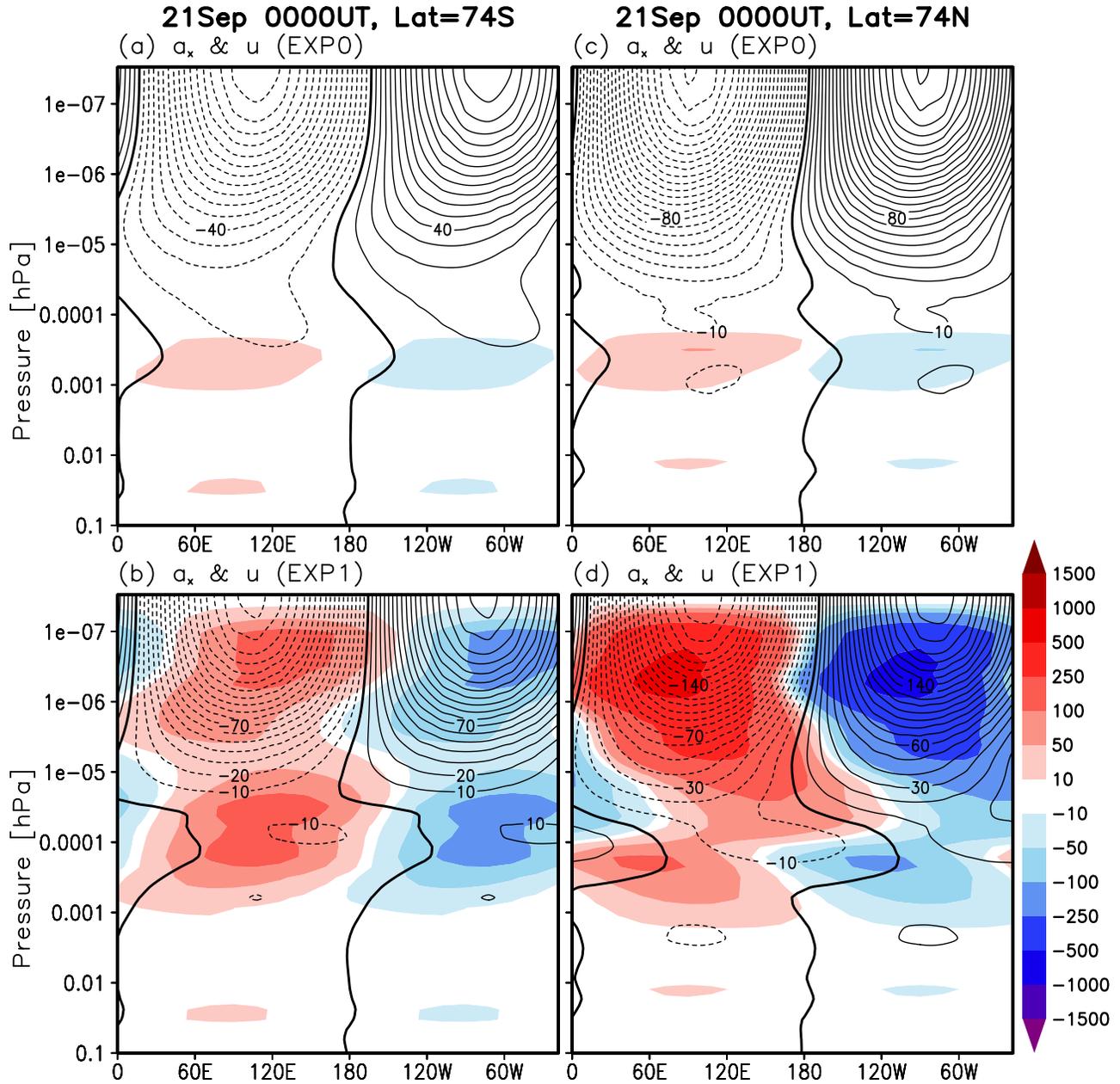}%
  \caption{Diurnal component of the zonal wind and the zonal GW drag presented in Figure
    \ref{fig:u_ax_wind_p_lon_highlat} at Southern (left) and Northern (right) Hemisphere
    high-latitudes on 21 Sept 0000 UT.}
  \label{fig:u_ax_wind_p_lon_highlat_diurnal}
\end{figure*}

At high-latitudes, thermospheric GW effects are more variable owing to additional energy and
momentum sources of magnetospheric origin that modulate the background
atmosphere. \textbf{Figure}~\ref{fig:u_ax_wind_p_lon_highlat} presents the results of
simulations for the Southern ($74^\circ$S) and Northern Hemisphere ($74^\circ$N)
high-latitudes. The varying depth of GW penetration into the upper thermosphere in the
different hemispheres is clearly visible. In the Southern Hemisphere, GWs saturate at lower
altitudes, depending on the longitude, and show a mixed action on the tide, partially (weakly)
accelerating or decelerating it, up to $\pm 250 $ m s$^{-1}$ day$^{-1}$, depending on the
phase: accelerative at 0--60$^{\circ}$E and 60$^\circ$E--120$^\circ$W; and decelerative at
120$^\circ$W--60$^\circ$W.  In the Northern Hemisphere, GWs penetrate to the upper
thermospheric altitudes, systematically imposing a substantial damping on the diurnal
variations of the zonal wind during both phases: 1500 m~s$^{-1}$ day$^{-1}$ during the easterly
phase and more than --250 m~s$^{-1}$ day$^{-1}$ peaking in the middle thermosphere during the
westerly phase. Note also that the inclusion of GW drag above the turbopause has a profound
effect on the simulated tide in the MLT below. In the EXP0 simulation
(\textbf{Figure}~\ref{fig:u_ax_wind_p_lon_highlat}a), the tidal variations of the zonal wind
are much weaker than in the EXP1 run (panel b).

An explicit analysis of DW1 signatures in the GW effects and in the tidal winds can provide
further insight into the GW forcing of the diurnal
tide. \textbf{Figures}~\ref{fig:u_ax_wind_p_lon_lowlat_diurnal} and
\ref{fig:u_ax_wind_p_lon_highlat_diurnal} display the longitudinal wavenumber-1 (diurnal)
components of the same quantities as they are presented at low-latitudes and high-latitudes in
Figures~\ref{fig:u_ax_wind_p_lon_lowlat} and \ref{fig:u_ax_wind_p_lon_highlat}. It is
immediately seen that (a) the longitudinal variations seen in
\textbf{Figures}~\ref{fig:u_ax_wind_p_lon_lowlat} are dominated by the wavenumber-1 component
(\textbf{Figure}~\ref{fig:u_ax_wind_p_lon_lowlat_diurnal}) associated with the DW1 tide and (b)
the diurnal component of the GW drag is in phase with the diurnal tide in the low-latitude MLT
in both simulations.  Higher in the middle and upper thermosphere, the correlation breaks down
and changes to anti-correlation, as noted above.  Longitudinal variations in the high-latitude
MLT behave differently: they are no longer composed of mainly the wavenumber-1 (DW1) component,
as is seen from \textbf{Figures}~\ref{fig:u_ax_wind_p_lon_highlat} and
\ref{fig:u_ax_wind_p_lon_highlat_diurnal}.  At high latitudes of both hemispheres, the diurnal
components of the GW forcing and wind oscillations are anti-correlated at all heights, although
it is seen that GWs tend to promote a development of the downward directed DW1 signature in the
MLT.

In order to understand these differences in the GW influence, we ought to reflect on the
physics of GW--tidal interactions. A wave harmonic with the horizontal phase velocity $c_i$ is
filtered at a critical level where $|c_i-\bar{u}| =0$.  Before this critical filtering
happens, i.e., at altitudes below the critical level (if the wave propagates from below), a
given harmonic attains an instability/saturation threshold at $|c_i-\bar{u}| \to |u'| \ne 0$,
where $|u'|$ is the wave amplitude of a single harmonic, or $|u'|=\sigma$ if the spectrum is
broad and consists of many harmonics $c_i$, $\sigma$ being the RMS wind fluctuations created
by all harmonics. Also, nonlinear dissipation ($\beta_{non}^i $) can significantly damp
  waves at lower altitudes, below the conventionally assumed linear instability level. The
surviving harmonics propagate higher, eventually either hitting their breaking/saturation
levels, or being obliterated by rapidly growing molecular diffusion and thermal conduction in
the upper thermosphere.

If the spectrum of GWs is sufficiently broad and propagates through alternating winds
associated with the tide, the momentum they deposit upon breaking/saturation tends to
accelerate the wind flow locally, thus maintaining and/or amplifying the tidal oscillations. In
the low-latitude stratosphere, GW harmonics generated at the tropopause encounter relatively
weak zonal mean winds in the stratosphere (\textbf{Figure}~\ref{fig:mean}), and many of them
avoid being filtered before arriving at the MLT. There, the harmonics obliterate while locally
amplifying the tide, as described above. The remaining waves propagate higher into the middle
and upper thermosphere, where strong in-situ generated DW1 tides dominate
(\textbf{Figure}~\ref{fig:u_ax_wind_p_lon_lowlat_diurnal}). There, the mechanism of GW-tide
interactions is similar to that of GWs with the mean zonal flow: because horizontal phase
velocities $c_i$ of the surviving harmonics are mostly slower or directed against the mean
wind, the main effect they produce is to drag the flow. At high-latitudes, the stratospheric
jets filter out a significant portion of the incident GW spectra. In addition, the wavenumber-1
component of the winds is contaminated by processes other than the DW1 tide. Therefore, the
effect of GWs on the tide in the high-latitude MLT is less certain and mixed. Higher in the
middle and upper thermosphere in the Northern Hemisphere, GWs tend to decelerate the in-situ
generated tide. In the Southern Hemisphere high-latitude, GWs provide a mixed effect on the
tide, leading overall to an enhancement of the tidal amplitude, which somewhat depends on the
variability of the high-latitude energy and momentum sources.

\section{Discussion}
\label{sec:discussion}
We presented the results of simulations focusing on the direct effects of parameterized
small-scale GWs on the diurnal migrating tides from the mesosphere to the upper thermosphere,
i.e., up to F-region altitudes. In this section, we discuss our results in the context of
previous observations and modeling studies for the MLT and upper thermosphere.

\subsection{Diurnal Migrating Tide and Gravity Wave Effects in the Mesosphere and Lower
  Thermosphere}
\label{sec:diurn-migr-tide_mlt}

Observational studies of the structure and seasonal behavior of the diurnal migrating tides in
the MLT are numerous \citep[e.g.,][]{Wu_etal06,Wu_etal08,Lieberman_etal10,AgnerLiu15}. For
example, \citet{Manson_etal02c} analyzed the data from the High Resolution Doppler Imager
(HRDI) onboard the Upper Atmosphere Research Satellite (UARS) for a two-month period of
September-October the global distribution of the migrating diurnal tidal winds at 96 km. They
found that the tidal amplitudes peak around $\pm 20-25^\circ$ latitude. Based on the
measurements with the WINDII (Wind Imaging Interferometer) instrument onboard UARS,
\citet{McLandress_etal96b} inferred the peak monthly mean amplitudes of 55--60 m~s$^{-1}$ for
the meridional wind component of DW1 tide located around $\pm20^\circ$. For the zonal wind
component, they found an asymmetry with respect to the equator with 30 m~s$^{-1}$ and 40
m~s$^{-1}$ in the Northern and Southern hemispheres, respectively. Our simulations
(\textbf{Figure}~\ref{fig:tideamp}) are in a good agreement with these observations, although
they slightly underestimate tidal amplitudes for the meridional wind. A closer examination
shows that the interhemispheric asymmetry in zonal winds and a symmetry for meridional winds
are better reproduced in the extended simulation (EXP1).

Our results are in a good qualitative agreement with the high-resolution simulations of
\citet{WatanabeMiyahara09}, who also suggested that GWs enhance the amplitude of the diurnal
migrating tide in the MLT. Although the model top in their study extended up to 150 km, the
conclusions concerning the impact of GWs on tides are applicable only up to $\sim$110-120 km,
because of the strong hyperdiffusion above the turbopause that effectively removed most of the
small-scale GWs.  Our study though includes small-scale effects in the whole atmosphere system
without any artificial diffusivity in the GCM.  However, their estimate of the diurnal
component of the zonal GW forcing in the low-latitude MLT ($\sim$8--15 m~s$^{-1}$~d$^{-1}$) is
about ten times smaller than in our simulations.  There are several possible reasons for this
discrepancy: \citet{WatanabeMiyahara09} (1) performed a perpetual run, while we conducted
day-stepping simulations, in which the GCM evolved in time, and (2) the employed horizontal
resolution (T213 or $\sim$0.5625$^\circ $) was apparently insufficient.  Our estimates for the
amplitude of the diurnal component of GW drag are close to $\sim$100 m~s$^{-1}$~d$^{-1}$ at
low-latitudes derived by \citet{LiuA_etal13} from meteor radar measurements.  Therefore,
\citet{WatanabeMiyahara09} have probably significantly underestimated GW diurnal variations and
small-scale GWs should be incorporated in modeling studies to better predict the variability
of the MLT. 

Recently, \citet{Gan_etal14} have used the extended CMAM model to examine the climatology of
the modeled migrating temperature tides. Our results are in very good agreement with their
tidal temperature amplitudes of $\sim$15 K centered around the equator in the MLT in
September. This comparison suggests that the extended simulation brings our results in better
agreement with the simulations of \cite{Gan_etal14} and the corresponding SABER data presented in
their paper.

There is an ongoing discussion on whether GWs increase the amplitude of DW1 in the MLT
\citep[e.g.,][]{England_etal06,LiuX_etal08}, decrease it \citep[e.g.,][]{Meyer99a}, or their
effects are mixed \cite[e.g.,][]{Lieberman_etal10,LiuA_etal13}.  Our simulations help to
explain the apparently contradicting results and to reconcile the views. As was demonstrated in
the previous section, a broad spectrum of GWs is required for enhancing tidal amplitudes. This
is why models utilizing GW parameterizations accounting for many harmonics tend to amplify the
tide, whereas those with few harmonics (Lindzen-type schemes) produce mainly damping. When the
tidal signature is weak and other processes participate in formation of the wind structure, the
influence of parameterized GWs is weak and uncertain.  Note that tidal variations and GW
momentum forcing may not necessarily be purely correlated or anticorrelated, but any phase
shift between them may occur. In such cases, GWs can amplify tide at one phase and damp it at
the other \citep{Lieberman_etal10}.

\subsection{Diurnal  Tide and Gravity Wave Effects in the Middle and Upper Thermosphere}
\label{sec:diurn-migr-tide_th}

The seasonal and latitudinal structures of the DW1 tide in the upper thermosphere as well as
its dependence on the solar activity are less studied observationally, even though the
associated amplitudes significantly exceed those in the MLT.  Numerical models have helped to
partially fill in this gap in the knowledge. Thermospheric tides have been simulated previously
with the first generation of thermospheric GCMs, which typically extended from the mesopause
region upward into the upper thermosphere \citep[e.g.,][]{Dickinson_etal81,
  Fesen_etal86,Fesen_etal93b}. These modeling efforts have demonstrated that the upper
thermospheric variability is dominated by the diurnal tide.  However, these models were not
designed to account for the lower atmospheric influence in a self-consistent manner.  To date,
no detailed study was performed to uncover the effects of GWs on the thermospheric tides with
the current generation of ``whole atmosphere" models.

Our simulations with controled propagation of parameterized GWs into the thermosphere
demonstrated that their influence on the tide is two-fold. The direct effect of subgrid-scale
GWs is to decrease the tidal amplitude in the Northern Hemisphere high-latitude, while they
increase the tidal amplitude in the low-latitude lower thermosphere. In the Southern Hemisphere
high-latitudes in the upper atmosphere, GWs modify the background atmosphere
and, thus, can ``indirectly" alter the structure and strength of the DW1 tide.  In
addition, GW drag can significantly modulate ion drag, thus contributing to the generation of
in-situ tides, mostly at high latitudes.

GW effects are continuously present in the both phases of the tide in the MLT. However, in the
upper thermosphere at high-latitudes depending on the wave dissipation, GWs may be present
during one phase of the tide and may have negligible effects during the opposite phase, as is
seen in the Southern Hemisphere, or may even be completely absent due to saturation at lower
altitudes. This should be realized when applying the space-time Fourier analysis to GW
dynamical and thermal effects in the thermosphere because GW signal in the thermosphere may not
be ``well-defined". Thus, it is sometimes instructive to study instantaneous distributions and
the mean fields to reveal the connection between thermospheric GW effects and tides, as was
done in our study.

\section{Summary and Conclusions}\label{sec:conc}
Using the Coupled Middle Atmosphere Thermosphere-2 General Circulation Model (GCM) with the
implemented nonlinear spectral gravity wave (GW) parameterization of \citet{Yigit_etal08}, we
explored the impact of small-scale GWs on the structure of the diurnal migrating tide (DW1) in
the middle and upper atmosphere during an equinox.  We presented the mean fields for the
September equinox conditions and quantified the GW dynamical and thermal effects on the general
circulation and temperature and on the amplitude of the
diurnal migrating tide. We performed this analysis by conducting (1) a cuttoff simulation, in
which GWs were removed above the turbopause in order to replicate previous modeling studies
that did not properly represent thermospheric GW effects, and (2) an extended
simulation, in which GWs were allowed to propagate self-consistently from their sources in the
lower atmosphere to the top of the model in the upper thermosphere. This approach helps 
isolate and identify the direct effects of small-scale GWs. The main findings of this study are
as follows:

  \begin{enumerate}
  \item The parameterized GWs affect the diurnal migrating tide directly and indirectly. Direct
    effects comprise a systematic phase shift between the GW forcing and tidal variations. The
    indirect effects include a substantial modification of the mean circulation induced by GWs,
    which affect the propagation of the DW1 component from the troposphere and its excitation
    in the middle and upper thermosphere.
  \item The simulated net effect of GWs on the diurnal migrating tide is to decrease or
    increase its amplitude in the thermosphere, depending on latitude.  In the low-latitude
    MLT, GWs enhance the tidal amplitudes. In the Northern Hemisphere high-latitude in the upper
    thermosphere, they damp the tides directly, while in the Southern Hemisphere
    high-latitudes, they indirectly lead to a tidal enhancement.
  \item In the low-latitude MLT, the correlation between the direction of the deposited GW
    momentum and tidal phase is caused by propagation of a broad spectrum of GW harmonics
    through the alternating winds. In the upper thermosphere and in middle- and high-latitudes
    of the MLT, mainly harmonics traveling against the local wind survive the selective
    filtering at the underlying levels. Thus, the deposited GW momentum imposes drag on the
    tidally-induced variations of the wind, similar to the well-studied effect on the zonal
    mean flow.
  \item These differences in influence of GWs on the migrating diurnal tide can be captured in
    GCMs if a GW parameterization (1) considers a broad spectrum of harmonics, (2) properly
    describes their propagation through a varying background, and (3) correctly accounts for
    the physics of wave breaking/saturation and, in particular, determines the altitude of
    harmonic's obliteration.
  \end{enumerate} 

Overall, GW activity and effects on the thermal tides are highly variable in the thermosphere
  due to the combined effects of lower atmospheric filtering, wave dissipation, ion-neutral
  coupling \citep{Yigit_etal09}, and the nonlinear response in the atmosphere.  Our study
  reveals the mechanics of GW--tidal interactions, while the actual effects that take place in
  the atmosphere must be studied observationally.

\acknowledgements{
EY has been partially funded by the NSF grant AGS 1452137. The modeling data
  supporting the figures presented in this paper can be obtained from EY
  (eyigit@gmu.edu).}

%
%

\end{document}